\def\year{2021}\relax
\documentclass[letterpaper]{article} %
\usepackage{aaai21}  %
\usepackage{times}  %
\usepackage{helvet} %
\usepackage{courier}  %
\usepackage[hyphens]{url}  %
\usepackage{graphicx} %
\def\UrlFont{\rm}  %
\usepackage{natbib}  %
\usepackage{caption} %
\title{Efficient On-Chip Learning for Optical Neural Networks Through Power-Aware Sparse Zeroth-Order Optimization}

\author {

       Jiaqi Gu\textsuperscript{\rm 1},
        Chenghao Feng \textsuperscript{\rm 1},
        Zheng Zhao \textsuperscript{\rm 2},
        Zhoufeng Ying \textsuperscript{\rm 3},
        Ray T. Chen \textsuperscript{\rm 1}, 
        David Z. Pan \textsuperscript{\rm 1}\\
}
\affiliations {
    \textsuperscript{\rm 1} University of Texas at Austin \\
    \textsuperscript{\rm 2} Synopsys, Inc.~
    \textsuperscript{\rm 3} Alpine Optoelectronics, Inc. \\
    \{jqgu, fengchenghao1996, zhengzhao, zfying\}@utexas.edu, \{chen, dpan\}@ece.utexas.edu
}
\usepackage[subrefformat=parens,farskip=0pt,justification=centering]{subfig}
\usepackage{microtype}
\usepackage{graphicx}
\usepackage{booktabs} %
\usepackage{amsmath}
\usepackage{threeparttable}
\usepackage{multirow}
\usepackage{bm}
\usepackage{bbm}
\usepackage{algorithm}
\usepackage[noend]{algpseudocode}
\usepackage{amssymb}
\usepackage{amsfonts}

\DeclareMathOperator*{\argmin}{arg\,min}

\graphicspath{{./figs/}}
\begin{document}
\maketitle

\begin{abstract}
\label{abstract}
Optical neural networks (ONNs) have demonstrated record-breaking potential in high-performance neuromorphic computing due to their ultra-high execution speed and low energy consumption.
However, current learning protocols fail to provide scalable and efficient solutions to photonic circuit optimization in practical applications.
In this work, we propose a novel on-chip learning framework to release the full potential of ONNs for power-efficient \textit{in situ} training.
Instead of deploying implementation-costly back-propagation, we directly optimize the device configurations with computation budgets and power constraints.
We are the first to model the ONN on-chip learning as a resource-constrained stochastic noisy zeroth-order optimization problem, and propose a novel mixed-training strategy with two-level sparsity and power-aware dynamic pruning to offer a scalable on-chip training solution in practical ONN deployment.
Compared with previous methods, we are the first to optimize over 2,500 optical components on chip.
We can achieve much better optimization stability, 3.7$\times$-7.6$\times$ higher efficiency, and save $>$90\% power under practical device variations and thermal crosstalk.
\end{abstract}

\section{Introduction}
\label{sec:Introduction}
As Moore's Law slows down, it becomes challenging for traditional electronics to further satisfy the escalating computational demands of machine learning tasks given clock frequency limitation and power density constraints.
Recently, the emerging optical neural network (ONN) has attracted increasing attention due to its ultra-high execution speed and order-of-magnitude higher energy efficiency compared to electronics.
In resource-constrained applications, ONNs become a promising alternative to accelerate machine learning workloads~\cite{NP_NATURE2017_Shen, NP_HotChips2020_Ramey, NP_NatureComm2020_Ying, NP_NanoPhotonics2020_Feng, NP_IEEEPJ2020_Ying}.
Computationally-intensive operations in neural networks, e.g., matrix multiplication, can be finished within the light propagation delay in one shot~\cite{NP_NATURE2017_Shen, NP_ASPDAC2019_Zhao, NP_ASPDAC2020_Gu, NP_TCAD2020_Gu, NP_CLEO2020_Feng, NP_APR2020_Miscuglio, NP_DATE2019_Liu, NP_DATE2020_Zokaee, NP_ICCAD2019_Zhao}.
With optical interconnects to reduce communication and memory transaction cost, a fully-optical neural engine provides a fundamental solution to break through the NN performance bound.
Shen, \textit{et al.}~\cite{NP_NATURE2017_Shen} demonstrated an integrated fully-optical neural chip to implement a multi-layer perceptron based on singular value decomposition (SVD)~\cite{NP_PHYSICAL1994_Reck,NP_OPTICA2016_Ribeiro}.
The weight matrices are mapped onto cascaded Mach-Zehnder interferometer (MZI) meshes to realize ultra-fast neural computing with over 100 GHz photo-detection rate and near-zero energy consumption~\cite{NP_NATURE2017_Shen,NP_OPTEXPRESS2012_Vivien}.

However, training methodologies for integrated ONNs still lack a scalable and efficient solution so far.
The mainstream approach offloads the training and simulation process to electrical computers using classical back-propagation (BP)~\cite{NP_NATURE2017_Shen, NP_ASPDAC2019_Zhao}, which is inefficient in circuit simulation and inaccurate in device noise modeling.
Hence, there exist great potentials to offload the learning process on photonic circuits.
Back-propagation is technically challenging to be implemented on a chip given the expensive hardware overhead and time-consuming gradient computation.

A brute-force phase tuning algorithm is proposed and adopted in~\cite{NP_NATURE2017_Shen,NP_JSTQE2020_Zhou} to perform ONN on-chip training via sequential device tuning, which is intractable as circuits scale up.
To mitigate the inefficiency issue of the above brute-force algorithm, an \textit{in situ} adjoint variable method (AVM)~\cite{NP_OPTICA2018_Hughes} is applied to directly compute the gradient w.r.t. MZI phases via inverse design.
However, it is challenging to be scaled to larger systems as the fully-observable circuits is a technically impractical assumption.
Evolutionary algorithms, e.g., genetic algorithm (GA) and particle swarm optimization (PSO), are introduced to train ONNs by population evolution~\cite{NP_OE2019_Zhang}.
A stochastic zeroth-order optimization framework \texttt{FLOPS}~\cite{NP_DAC2020_Gu} has been proposed to improve the ONN learning efficiency by 3-5 times via random-sampling-based zeroth-order gradient estimation.

However, previous works have the following disadvantages: 1) nontrivial Gaussian sampling cost, 2) divergence issues due to high variance, 3) high energy consumption, and 4) hardware-unfriendly weight update step size.
In this work, we propose a novel mixed-training framework that enables scalable on-chip optimization with more stable convergence, higher training efficiency, and much lower power consumption under non-ideal environment.
Compared with previous state-of-the-art (SOTA) methods, our mixed-training framework has the following advantages.
\begin{itemize}
    \item Efficiency: our mixed-training strategy achieves 3$\sim$7$\times$ fewer ONN forward and much lower computation complexity than SOTA ONN on-chip learning methods.
    \item Robustness: our method adopts a novel optical device mapping with mixed-active/passive regions to protect ONNs from device variations and thermal crosstalk, leading to better noise-tolerance than previous solutions.
    \item Stability: our stochastic zeroth-order sparse coordinate descent optimizer (\texttt{SZO-SCD}) outperforms SOTA zeroth-order optimizers with more stable convergence and better performance in on-chip accuracy recovery.
    \item Scalability: our proposed optimizer leverages two-level sparsity in on-chip training, extending the ONN learning scale to $>$2,500 MZIs.
    \item Power: we propose a lightweight power-aware dynamic pruning technique, achieving $>$90\% lower power consumption with near-zero accuracy loss or computation overhead.
\end{itemize}

\section{Preliminary}
\label{sec:Preliminary}
In this section, we will introduce the architecture of integrated ONNs, prior work in ONN on-chip training, and background knowledge about stochastic zeroth-order optimization.
\subsection{ONN Architecture and Training Methods}
\label{sec:ONNPrinciple}
\begin{figure}
\vspace{-5pt}
    \centering
    \includegraphics[width=0.45\textwidth]{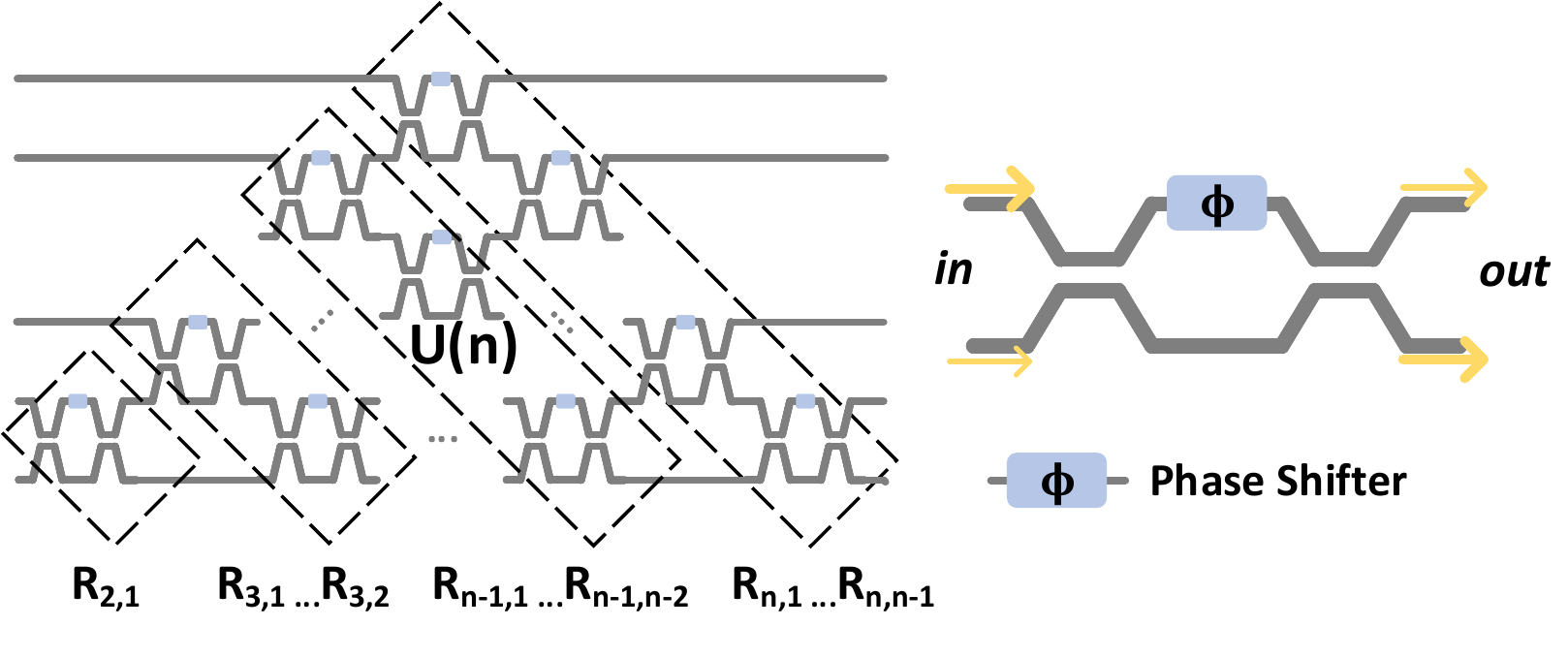}
    \vspace{-8pt}
    \caption{MZI triangular array $\bm{U}(n)$ and the MZI structure.}
    \label{fig:MZIArray}
    \vspace{-15pt}
\end{figure}
The integrated optical neural network (ONN) is a hardware platform that implements artificial neural networks with silicon-photonics.
As a case study, we focus on an ONN architecture based on singular value decomposition (SVD)~\cite{NP_NATURE2017_Shen}.
It decomposes an $m \times n$ weight matrix using SVD, i.e., $\bm{W}=\bm{U\Sigma V^{\ast}}$.
The diagonal matrix $\bm{\Sigma}$ can be simply implemented by on-chip attenuators, e.g., single-port Mach-Zehnder interferometers (MZIs), to perform signal scaling.
The unitary matrices $\bm{U}$ and $\bm{V}^{\ast}$ can be realized by a cascaded MZI triangular array~\cite{NP_PHYSICAL1994_Reck},
shown in Fig.~\ref{fig:MZIArray}.
The unitary group parametrization is given by,
\begin{equation}
\small
\label{eq:UnitaryParametrization}
\bm{U}(n)=\bm{D}\;\prod_{i=n}^{2}\prod_{j=1}^{i-1}\bm{R}_{ij}(\phi_{ij}),
\end{equation}
where $\bm{D}$ is a diagonal matrix with $\pm{1}$ on its diagonal entries, and the 2-dimensional planar rotator $\bm{R}_{ij}(\phi_{ij})$ is an $n$-dimensional identity matrix where entries on ($i$,$i$), ($i$,$j$), ($j$,$i$), ($j$,$i$) are $\cos{\phi_{ij}}$, $\sin{\phi_{ij}}$, -$\sin{\phi_{ij}}$, $\cos{\phi_{ij}}$, respectively.
Each rotator $\bm{R}_{ij}$ can be implemented by a 2$\times$2 MZI that produces unitary interference of input light signals with a rotation angle $\phi$ as follows~\cite{NP_NATURE2017_Shen},
\begin{equation}
\small
    \begin{pmatrix}
    y_1 \\
    y_2
    \end{pmatrix}=
    \begin{pmatrix}
    \cos{\phi} & -\sin{\phi} \\
    \sin{\phi} & \cos{\phi}
    \end{pmatrix}
    \begin{pmatrix}
    x_1 \\
    x_2
    \end{pmatrix}.
\end{equation}

To train ONNs, the traditional procedure trains the weight matrix $\bm{W}$ using gradient back-propagation and then maps it to photonic circuits through SVD and unitary group parametrization~\cite{NP_PHYSICAL1994_Reck}, which is inefficient and hardware-agnostic.
Later, several ONN on-chip learning protocols are proposed to perform \textit{in situ} circuit optimization.
To solve the problem, a straightforward approach is to compute the gradient w.r.t each MZI configuration given by,
\begin{equation}
\small
\begin{aligned}
    \label{eq:GradientUnitaryParametrization}
    \frac{\partial\mathcal{L}}{\partial\bm{R}_{ij}}&=\big(\bm{D}\bm{R}_{n1}\bm{R}_{n2}\bm{R}_{n3}\big)^T\nabla_{y}\mathcal{L}~x^{T}\big(\cdots\bm{R}_{32}\bm{R}_{21}\bm{\Sigma}\bm{V}^{\ast}\big)^T\\
    \frac{\partial\mathcal{L}}{\partial\phi_{ij}}&=\mathrm{Tr}\bigg(\Big(\frac{\partial\mathcal{L}}{\partial\bm{R}_{ij}}\odot\frac{\partial{\bm{R}_{ij}}}{\partial{\phi_{ij}}}\Big)(e_i+e_j)(e_i+e_j)^T\bigg),
\end{aligned}
\end{equation}
where $e_i$ is the $i$-th orthonormal basis.
On edge computing platforms, this analytical Jacobian is computationally-prohibitive, especially since $\nabla_y\mathcal{L}~x^T$ is intractable in practical deployment.
Later, a brute-force phase tuning method is proposed~\cite{NP_NATURE2017_Shen, NP_JSTQE2020_Zhou} using finite-difference-based gradient estimation.
Adjoint variable method (AVM)~\cite{NP_OPTICA2018_Hughes} is proposed to model the circuit state as a partial-differential-equation-controlled linear system, and directly measures the exact gradient via \textit{in situ} light intensity measurement.
Evolutionary algorithms, e.g., particle swarm optimization and genetic algorithm, are demonstrated to train MZIs on chip~\cite{NP_OE2019_Zhang}.
A stochastic zeroth-order gradient descent based method \texttt{FLOPS}~\cite{NP_DAC2020_Gu} has been proposed to improve the training efficiency by 3-5$\times$ compared with previous methods.

\subsection{Stochastic Zeroth-Order Optimization}
\label{sec:ZOO}
To solve optimization problems when analytical Jacobian is infeasible to compute, zeroth-order optimization (ZOO) plays an significant role, e.g., black-box adversarial attacks, policy-gradient-based reinforcement learning, and circuit parameter optimization~\cite{NN_AIsec2017_Chen,NN_NIPS2019_Chen,NN_SIAM2013_Saeed,NN_NIPS2018_Liu,NN_ICLR2020_Gorbunov,NN_AAAI2020_Zhao,NN_AAAI2019_Tu,NN_AAAI2018_Wang}.
Various ZOO methods have been proposed with mathematically-proven convergence rate, including stochastic gradient descent with Nesterov's acceleration~\cite{NN_FCM2017_Nesterov}, zeroth-order coordinate-wise Adam \texttt{ZOO-ADAM} and Newton's method \texttt{ZOO-Newton}~\cite{NN_AIsec2017_Chen}, zeroth-order adaptive momentum method \texttt{ZOO-AdaMM}~\cite{NN_AIsec2017_Chen}, stochastic three-points~\cite{NN_AAAI2020_Bibi}, stochastic momentum three-points~\cite{NN_ICLR2020_Gorbunov}, etc.
Most zeroth-order optimizers have a convergence rate dependent on the dimensionality, which intrinsically makes them less efficient and less scalable than higher-order optimizers.
In this work, we explore two-level sparsity in stochastic zeroth-order optimization to enable scalable, stable, and efficient ONN on-chip training.
\begin{figure*}
    \centering
    \includegraphics[width=0.8\textwidth]{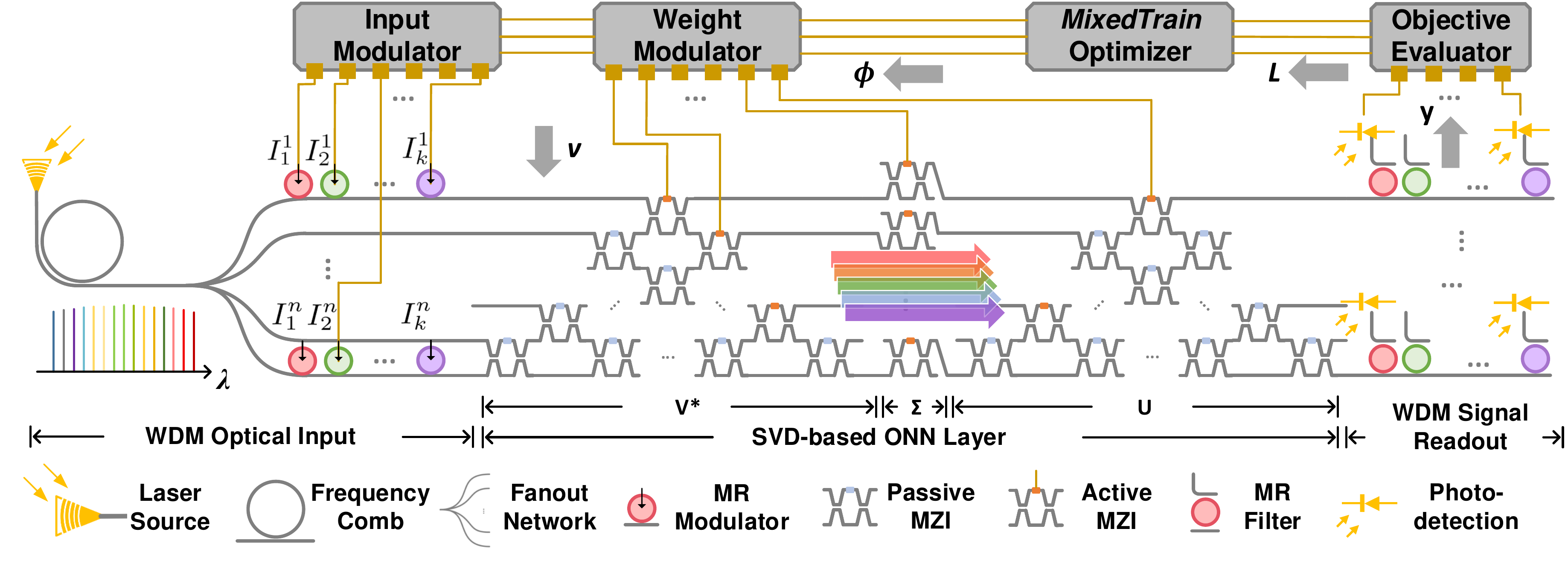}
    \vspace{-7pt}
    \caption{Schematic of ONN on-chip learning framework with stochastic zeroth-order mixed-training.}
    \label{fig:FLOPSFramework}
    \vspace{-10pt}
\end{figure*}

\section{Problem Formulation and Analysis}
\label{sec:ProblemFormulation}
Before discussing our proposed on-chip learning framework, we give a formulation to the resource-limited ONN learning problem.
In practical ONN applications, the ultimate target is to leverage photonic neural chip to complete machine learning tasks with high accuracy, low latency, and low energy consumption, under environmental changes and device-level variations.
The optimization variables are optical device configurations, i.e., phase shift for all MZIs $\bm{\Phi}$, including those in the unitary matrices $\bm{U}$ and $\bm{V}^{\ast}$ and the diagonal matrix $\bm{\Sigma}$.
The objective is the task-specific loss function.
We are the first to formulate the ONN on-chip learning as a resource-limited accuracy recovery problem in the unitary space,
\begin{align}
    \small
    \label{eq:Formulation}
    &\bm{\Phi}^{\ast}=\argmin_{\bm{\Phi}\sim\mathcal{R}}\mathcal{L}_{\mathcal{D}^{trn}}(\bm{\bm{W}(\bm{\Phi})}),\\
     \text{s.t.} ~~ &\bm{W}(\bm{\Phi})=\bm{U}(\bm{\Phi}^U)\bm{\Sigma}(\bm{\Phi^S}) \bm{V^{\ast}}(\bm{\Phi}^V), \notag \\
     &\bm{U}(\bm{\Phi}^U)=\bm{D}^U\;\prod_{i=N}^{2}\prod_{j=1}^{i-1}\bm{R}_{ij}(\phi^U_{ij}),\notag\\
     &\bm{V}^{\ast}(\bm{\Phi}^V)=\bm{D}^V\;\prod_{i=M}^{2}\prod_{j=1}^{i-1}\bm{R}_{ij}(\phi^V_{ij}),\notag\\
     &\|\bm{\Sigma}(\bm{\Phi}^S)\|_{\infty} < m, \notag\\
     &\bm{\Phi}\in[0, 2\pi), \notag\\
     &\texttt{Power}(\bm{\Phi}) \leq \tilde{P}, \int_t\texttt{Power}(\bm{\Phi}^t)\;\text{d}t \leq \tilde{E}, \notag\\
     &\mathcal{C}(\nabla_{\bm{\Phi}}\mathcal{L})\gg\tilde{C}\gg\mathcal{C}(\mathcal{L}), \mathcal{C} \leq \tilde{C},\notag
\end{align}
where $\mathcal{D}^{trn}$ is the training set.
In each layer, the weight matrix $\bm{W}\in\mathbb{R}^{M\times N}$ is constructed by $\bm{U}$, $\bm{\Sigma}$, and $\bm{V}^{\ast}$, where $\bm{U}$ and $\bm{V}^{\ast}$ are constrained in the Stiefel manifold, and the $\ell_{\infty}$-norm of the diagonal matrix $\bm{\Sigma}$ is bounded by a empirically largest signal scaling range $m$.
The optimization parameters $\bm{\Phi}$ are constrained in a hypercube within 0 degree and $2\pi$ degree.
The photonic device programming power has to honor a maximum power budget $\tilde{P}$ during ONN inference.
Also, the total energy used to program ONN devices during on-chip training is bounded by an energy budget $\tilde{E}$.
The last constraint is the computation budget for the optimizer, which can not afford to calculate the Jacobian $\nabla_{\bm{\Phi}}(\mathcal{L})$ shown in Eq.~\eqref{eq:GradientUnitaryParametrization}, but the objective evaluation is ultra-fast with optics.

To solve the optimization problem on this SVD-based architecture, we directly optimize the decomposed matrices $\bm{U}$ and $\bm{V}^{\ast}$ within the Stiefel manifold.
Previous work proposed Riemannian optimization~\cite{NN_AAAI2018_Huang}, unitary regularization~\cite{NP_ASPDAC2019_Zhao}, and unitary projection~\cite{NP_DATE2020_Gu} to satisfy the unitary constraints.
In this work, we optimize the phases $\bm{\Phi}$ in the Reck-style unitary parametrization space to achieve minimum computation complexity.
For the diagonal matrix, we optimize it as $\texttt{diag}(\bm{\Sigma})=m(\cos{\phi^S_0},\cdots,\cos{\phi^S_{\min(M,N)-1}})$, such that the optimization variables can be unified with $\bm{\Phi}^U$ and $\bm{\Phi}^V$.
To facilitate power optimization, we do not use the periodic phase space relaxation~\cite{NP_DAC2020_Gu}.
Instead, we wrap the phase within the valid hypercube by $\phi=(\phi\mod 2\pi)$ at each iteration avoid unnecessary power once the updated phase exceeds $2\pi$.
To meet the computation budget, we will introduce a lightweight technique to handle power and energy constraints.
The computation budget of the resource-constrained platform can be satisfied by manual searching in the optimizer design space, where lightweight zeroth-order optimization methods will be promising candidates.

\section{Proposed ONN On-Chip Learning Framework}
\label{sec:ProposedFramework}
\begin{figure*}
    \centering
    \includegraphics[width=0.98\textwidth]{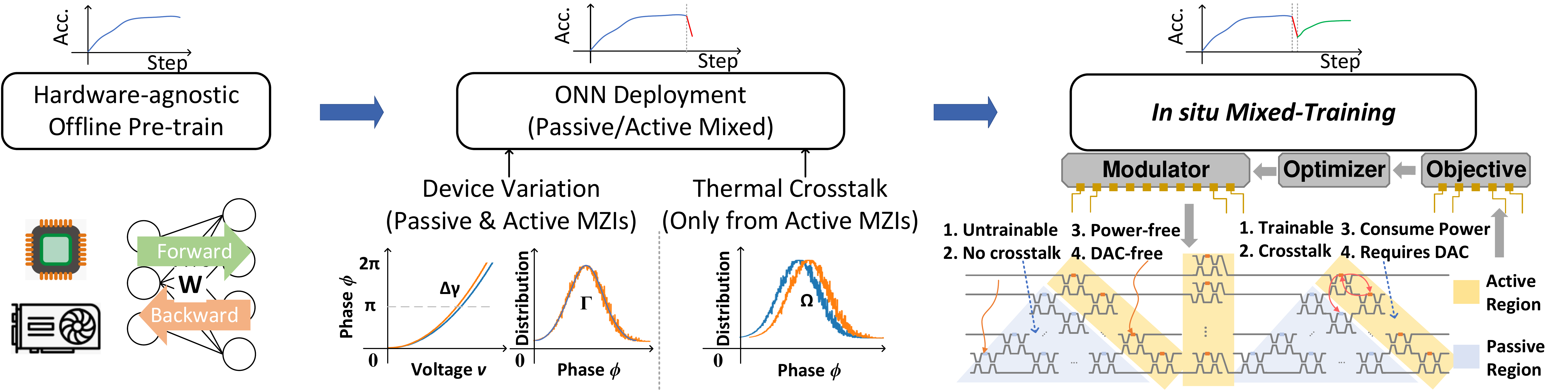}
    \caption{Mixed-training flow in the practical ONN deployment.}
    \label{fig:MixedTrainFlow}
    \vspace{-5pt}
\end{figure*}

In the practical ONN deployment, apart from the basic constraints listed in Eq.~\eqref{eq:Formulation}, non-ideal environment and device noises are necessary to be considered in the learning framework.
Therefore, we propose a mixed-training strategy to efficiently solve this noisy learning problem with all the aforementioned constraints in Fig.~\ref{fig:MixedTrainFlow}.

\subsection{Scalable Mixed-Training Strategy}
\label{sec:MixedTraining}
To enable efficient ONN on-chip learning on practical networks and dataset, we propose a mixed-training strategy to reduce the optimization dimensionality and minimize tunable devices for better convergence and lower power consumption.
Specifically, we assume a model is pre-trained and prepared for edge ONN deployment.
Then, our target is to implement the pre-trained model on practical ONN engines while recovering the accuracy given non-ideal environment and device variations.
The naive solution is to deploy a fully active ONN where all optical devices are thermo-optically tunable with maximum learnability~\cite{NP_DAC2020_Gu}.
However, This leads to high control complexity, power consumption, and non-ideal thermal crosstalk.
In this work, we propose a mixed-training strategy that integrates passive and active ONNs to balance efficiency, robustness, and learnability, shown in Fig.~\ref{fig:MixedTrainFlow}.
Now we introduce three stages in the entire ONN on-chip mixed training flow.
\subsubsection{Hardware-unaware Pre-training}
\label{sec:Pretrain}
For the MZI-based ONN architecture,
Hardware-unaware training based on back-propagation is firstly performed with an ideal computational model on electrical digital platforms, e.g., GPUs and CPUs, to obtain target device configurations.

\subsubsection{ONN Deployment with Mixed Active/Passive Regions}
We proposed to deploy the ideally-trained model on the photonic circuits using a mixed passive/active design, where most parameters in two unitary matrices $\bm{U}$ and $\bm{V}^*$ are fixed by using passive optical devices.
Only the diagonal matrix $\bm{\Sigma}$ and a small fraction ($\alpha\ll 1$) of phases in two unitary matrices are implemented by active devices to enable its adaptability and learnability.
We denote fixed phases in the passive region as $\mathcal{P}$ and tunable phases in the active region as $\mathcal{A}$.

In practical applications, device variations, e.g., phase shifter $\gamma$ coefficient drift, and thermal crosstalk among MZIs will be present, leading to output perturbation and thus accuracy loss, shown in Fig.~\ref{fig:MixedTrainFlow}.
The phase shifter variations come from environmental temperature changes or manufacturing errors.
Under variations, the power-to-phase-shift factor $\gamma$ of both active and passive phase shifters will drift from the ideal value as $\gamma^v=\gamma+\Delta\gamma$, where we assume the noise is sampled from a truncated Gaussian distribution $\Delta\gamma\in\mathcal{N}(0,\sigma_{\gamma}^2)$.
We assume the rotation angle is proportional to the device-related coefficient as $\phi\propto\gamma$, the noisy phase is denoted as $\phi^v=\phi\gamma^v/\gamma$.
For $N$ rotation angles, this variation is described as a diagonal perturbation matrix $\bm{\Phi^v}=\bm{\Gamma}\bm{\Phi}$.
In terms of thermal crosstalk, the correlated heat distribution among thermo-optic devices leads to an increase in the steady temperature.
In the heat steady state, the mutual correlation of phases within $N$ noisy rotation angles $\bm{\Phi}^v$ can be described by a coupling matrix as $\bm{\Phi}^c=\bm{\Omega}\bm{\Phi^v}$,
{\small\begin{align}
    \small\centering
    \label{eq:Crosstalk}
    \begin{pmatrix}
    \phi^c_0\\
    \phi^c_1\\
    \vdots\\
    \phi^c_{N-1}
    \end{pmatrix}=&
    \begin{pmatrix}
    \omega_{0,0} & \omega_{0,1} & \cdots & \omega_{0,N-1} \\
    \omega_{1,0} & \omega_{1,1} & \cdots & \omega_{1,N-1} \\
    \vdots & \vdots & \ddots & \vdots \\
    \omega_{N-1,0} & \omega_{N-1,1} & \cdots & \omega_{N-1,N-1}
    \end{pmatrix}
    \begin{pmatrix}
    \phi^v_0\\
    \phi^v_1\\
    \vdots\\
    \phi^v_{N-1}
    \end{pmatrix}\notag\\
    \rm{s.t.}~~&\omega_{i,j}=1, \quad \forall\;i=j\notag\\
    &\omega_{i,j}=0, \quad \forall\;i\neq j \text{ and }\phi_j \in \mathcal{P}\\
    &0\leq\omega_{i,j}<1, \quad \forall\;i\neq j \text{ and }\phi_j \in \mathcal{A}\notag.
\end{align}}%
The diagonal factor $\omega_{i,j}, i=j$ is the self-coupling coefficient, which is typically set to 1.
$\omega_{i,j}, i\neq j$ is the mutual coupling coefficient~\cite{NP_JLT2019_milanizadeh}.
As a physical fact, only active devices are thermal aggressors that perturb adjacent devices, while passive devices do not impact their neighbors since they have zero heat dissipation.
Hence mutual coupling factors $\omega_{i,j}, i\neq j$ are set to 0 if $\phi_j$ represents a passive MZI.
We can unify the $\gamma$ noise with the crosstalk as $\bm{\Phi^c}=\bm{\Omega}\bm{\Gamma}\bm{\Phi}$.
Therefore, the objective is re-written as,
\begin{equation}
    \small
    \label{eq:FormulationNoisy}
    \bm{\Phi}^{\ast}=\argmin_{\bm{\Phi}\sim\mathcal{R}}\mathcal{L}_{\mathcal{D}^{trn}}(\bm{\bm{W}(\bm{\Omega\Gamma\Phi})}).
\end{equation}

\subsubsection{Mixed-training with Stochastic Zeroth-Order Sparse Coordinate Descent (SZO-SCD)}
\label{sec:MixedTrainWithSZO-SCD}
In this stage, we introduce stochastic zeroth-order sparse coordinate descent (\texttt{SZO-SCD}) to tune the active devices for \textit{in situ} accuracy recovery.
Since the pre-trained model is roughly converged, the ZO-gradient-based method~\cite{NP_DAC2020_Gu} will suffer from divergence issues due to its gradient estimation variance.
In contrast, our \texttt{SZO-SCD} optimizer is more suitable for near-convergence fine-tuning in the phase space.
In iteration $t$, only a fraction ($s \ll 1$) of active devices $\bm{\Phi}_s=\{\phi_0,\cdots,\phi_{s|\mathcal{A}|-1}\}\subseteq\mathcal{A}$ are selected for coordinate descent as follows,
\begin{equation}
    \small
    \label{eq:SZO-SCD}
    \phi^{t+1}_i\gets\argmin_{\phi_i}\{\mathcal{L}_{\mathcal{I}^t}(\phi^t_i+\delta\phi), \mathcal{L}_{\mathcal{I}^t}(\phi^t_i-\delta\phi)\}.
\end{equation}
The mini-batch evaluation of $\mathcal{L}_{\mathcal{I}}(\cdot)$ can be processed in parallel by using WDM~\cite{NP_OE2014_Tan, NP_PW2020_Feng} shown in Fig.~\ref{fig:FLOPSFramework}.

The advantages of the mixed-training strategy with \texttt{SZO-SCD} lie in several aspects.
First, since the method reduces the tunable parameters of an $N\times N$ weight matrix from $N^2$ to $N + s\alpha N^2$ per iteration, the optimization efficiency will be considerably improved.
Second, the passive ONN part consumes nearly zero energy, leading to approximately (1-$\alpha$) power saving.
Third, the optimization dimensionality is reduced from the full $\mathcal{O}(N^2)$ space to a sparse subspace, which accelerates the convergence of our zeroth-order learning algorithm with a slimmed computation demand.
Fourth, this method has $\mathcal{O}(1)$ computation complexity and $\mathcal{O}(1)$ memory complexity per iteration, which is nearly the cheapest optimizer in the design space.

\subsection{Power-aware Dynamic Pruning}
\label{sec:GradientPruning}
On resource-limited edge applications, low power consumption will be a preferable feature to enhance endurance.
We assume the power of active phase shifter is proportional to the rotation angle $P\propto\phi$, then we can use the phase $\phi\in[0, 2\pi)$ as a fast device tuning power estimator.
A straightforward approach to handle this power constraint is to use Lagrangian relaxation to add the power constraint in the objective as follows,
\begin{equation}
    \small
    \label{sec:PowerRelaxation}
    \begin{aligned}
    \bm{\Phi}^{\ast}=&\argmin_{\bm{\Phi}\sim\mathcal{R}}\mathcal{L}_{\mathcal{D}^{trn}}(\bm{W}(\bm{\Omega\Gamma\Phi})) + \lambda P(\bm{\Phi)}\\
    P(\bm{\Phi})=&\sum_{\phi\in\bm{\Phi}}(\bm{\phi}\!\!\!\mod 2\pi),
    \end{aligned}
\end{equation}
and solve it using alternating direction multiplier method (ADMM).
However, the dual update for power optimization will cause convergence issues, which will be shown in our later experiment.
To implicitly consider power constraints in the optimization, we propose a power-aware dynamic pruning technique to further boost the power efficiency with stable convergence.
The detailed power-aware optimization algorithm is described in Alg.~\ref{alg:PowerAwareSZO-SCD}.
\begin{algorithm}[tb]
\caption{\texttt{SZO-SCD} with Power-aware Dynamic Pruning}
\label{alg:PowerAwareSZO-SCD}
\begin{algorithmic}[1]
\small
\Require ONN forward function $\mathcal{L}(\cdot)$, phases $\bm{\Phi}^0$ after ONN deployment, training dataset $\mathcal{D}^{trn}$, total iterations $T$, Active set $\mathcal{A}$,
sparsity of fine-tuned phases $s$, initial tuning step size $\delta\phi^0>0$, and power awareness $p\in[0,1]$, power estimator \texttt{power}($\cdot$);
\Ensure Converged phases $\bm{\Phi}^{T-1}$;

\For{$t\gets 0\cdots T-1$}
\State Randomly sample a mini-batch $\mathcal{I}^t$ from $\mathcal{D}^{trn}$
\State Randomly select $\bm{\Phi}^t_s=\{\phi^t_0,\cdots,\phi^t_{s|\mathcal{A}|-1}\} \subseteq\mathcal{A}$ without replacement
\For {$\phi_i^t \gets \phi^t_0,\cdots,\phi^t_ {s|\mathcal{A}|-1}$}
\If{$\mathcal{L}_{\mathcal{I}^t}(\phi_i^t+\delta\phi^t)<\mathcal{L}_{\mathcal{I}^t}(\phi_i^t)$}
\State $\phi_i^{t+1} \gets \phi_i^t+\delta\phi^t$
\Else
\If{$\texttt{power}(\phi_i^t-\delta\phi^t)>\texttt{power}(\phi_i^t)$}
\State $b\sim\mathcal{B}(p)$ \Comment{Sample from Bernoulli distribution with probability $p$ to take 1}
\State $\phi_i^{t+1} \gets \phi_i^t-b\cdot\delta\phi^t$
\Else
\State $\phi_i^{t+1} \gets \phi_i^t-\delta\phi^t$
\EndIf
\EndIf
\EndFor
\State $\delta\phi^{t+1}=\texttt{Update}(\delta\phi^t$) \Comment{Step size decay}
\EndFor
\end{algorithmic}
\end{algorithm}
In lines 8-12, the optimizer will prune backward steps with probability $p$ if the objective increases in the positive direction and the power consumption increases in the negative direction.
The intuition behind this efficient power-aware pruning is that our \texttt{SZO-SCD} only queries the zeroth-order oracle, such that the step-back is not guaranteed to be a descent direction.
This uncertainty enables us to embed a power-constraint handling mechanism to dynamically prune step-backs that do not have a descent guarantee but lead to a certain power increase.
The probabilistic power-awareness factor $p$ also provides a parametric approach to balance between power and solution quality.

\vspace{-.05in}
\section{Experimental Results}
\label{sec:ExperimentalResults}
\begin{figure*}
    \centering
    \subfloat[]{\includegraphics[width=0.28\textwidth]{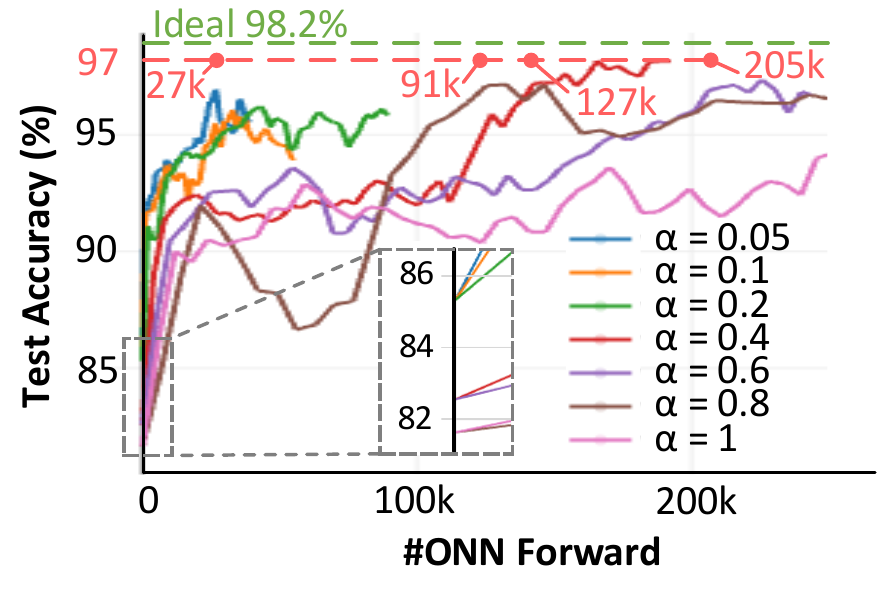}
    }
    \subfloat[]{\includegraphics[width=0.28\textwidth]{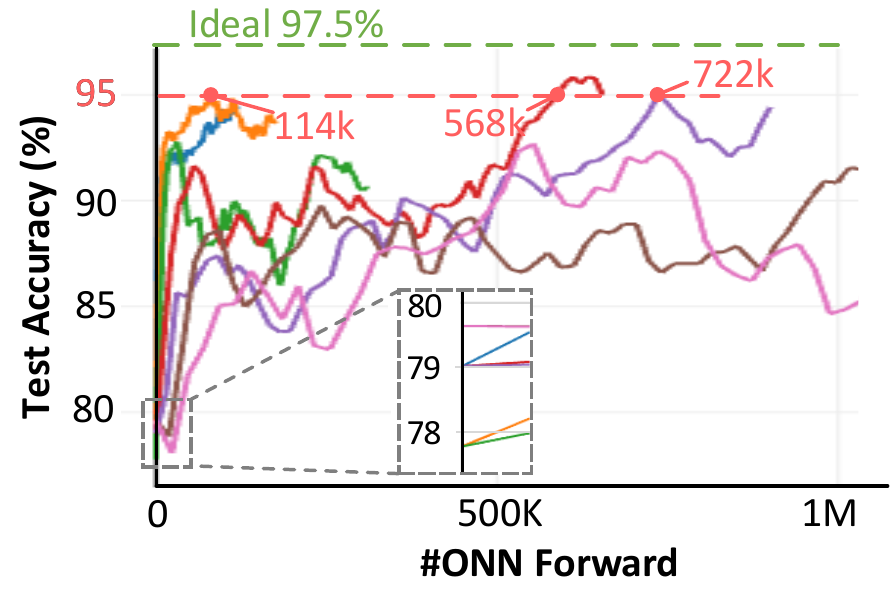}
    }
    \subfloat[]{\includegraphics[width=0.28\textwidth]{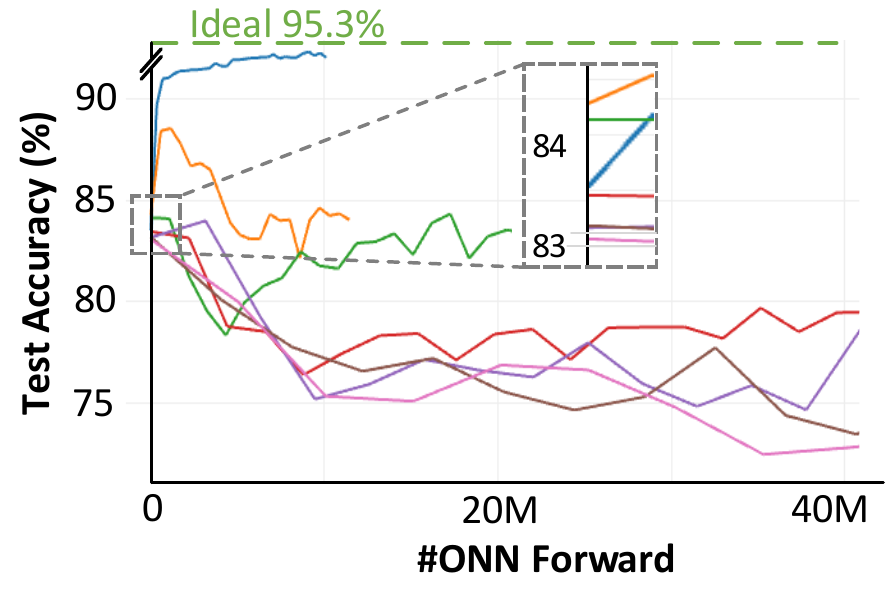}
    }
    \caption{Test accuracy with different mixed-training sparsity $\alpha$.
    (a) MLP (8-16-16-4) on Vowel Recognition, (b) MLP (10-24-24-6) on Vowel Recognition, and (c) MLP (8$\times$8-24-24-10) on MNIST.
    Close-up views show the accuracy after deployment.}
    \label{fig:MixedTrainEval}
    \vspace{-5pt}
\end{figure*}
\begin{figure*}
    \centering
    \subfloat[]{\includegraphics[width=0.28\textwidth]{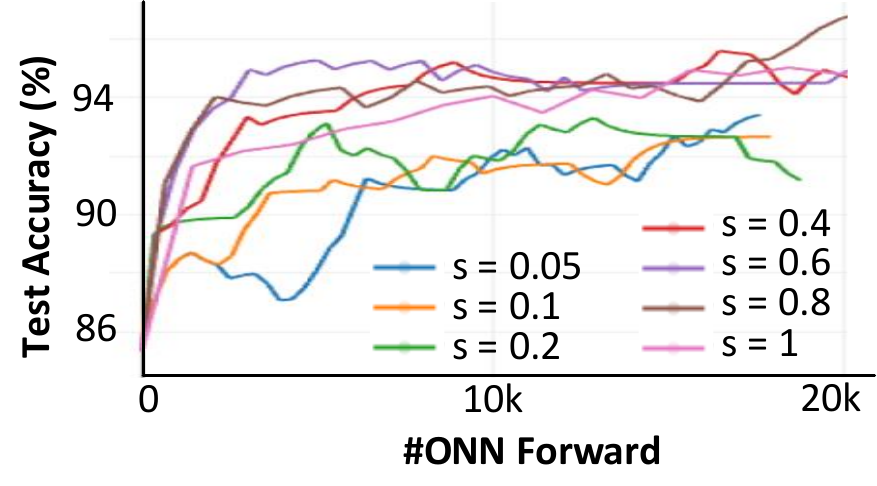}
    }
    \subfloat[]{\includegraphics[width=0.28\textwidth]{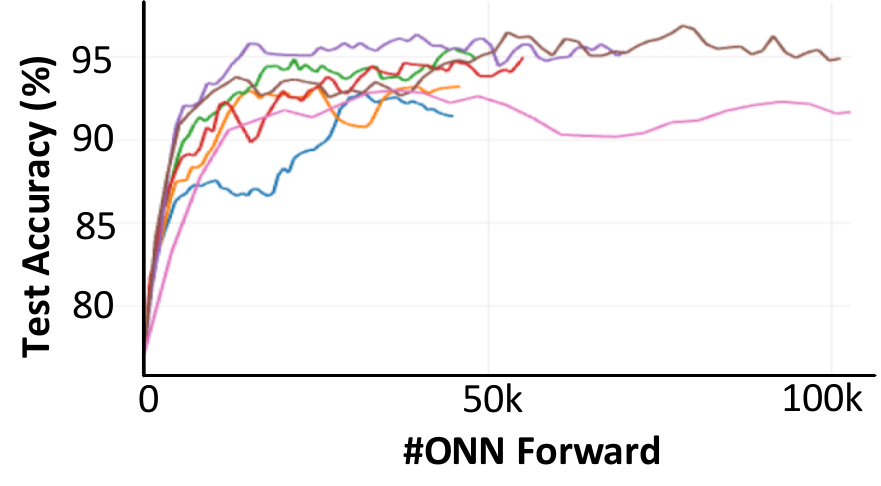}
    }
    \subfloat[]{\includegraphics[width=0.28\textwidth]{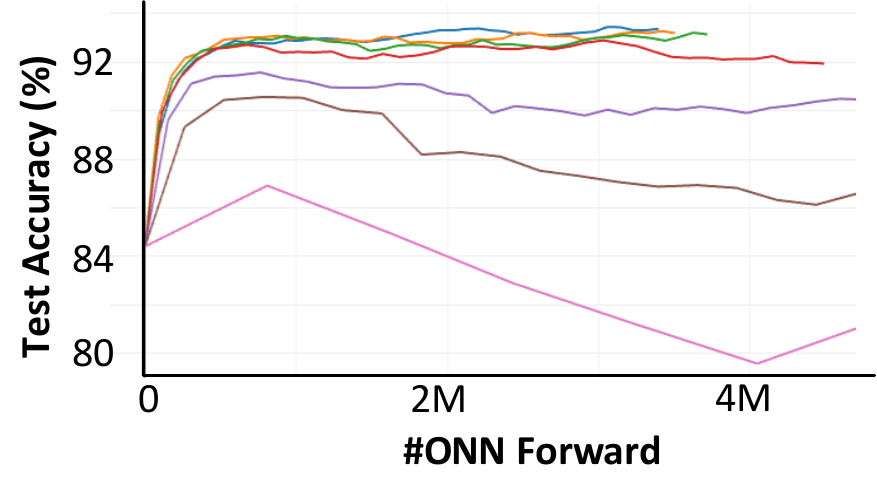}
    }
    \caption{Evaluation with different sparsity $s$ in \texttt{SZO-SCD}.
    $\alpha$ is set to 0.15 for all models.
    (a) 8-16-16-4 on Vowel Recognition dataset.
    (b) 10-24-24-6 on Vowel Recognition dataset.
    (c) (8$\times$8)-24-24-10 on MNIST dataset.
    }
    \label{fig:SparseTuneEval}
    \vspace{-5pt}
\end{figure*}
Experiments are conducted on a vowel classification dataset~\cite{NN_Vowel1989}, MNIST~\cite{NN_MNIST1998}, FashionMNIST~\cite{NN_FashionMNIST2017}, and CIFAR-10~\cite{NN_CIFAR2009} for image classification.
As a proof-of-concept demonstration, those datasets are standard and practical for ONNs.
We implement all methods in PyTorch with an NVIDIA Quadro RTX 6000 GPU and an Intel Core i7-9700 CPU.
We adopt a step size $\delta\phi$=0.02 with an epoch-wise exponential decaying rate of 0.985 and a mini-batch size of 32.
The upper bound $m$ set for $\bm{\Sigma}$ is 3.
Following a common setting, the std. of phase variation $\sigma(\gamma)$ is set to 2e-3 and the mutual-coupling factor $\omega$ is set to 2e-3 only for adjacent MZIs.
Rectified linear units (ReLU)~\cite{NN_ICCV2015_He} with an upper limit 4 are used as the nonlinearity.

\subsection{Evaluation on Mixed-Training Strategy}
\label{sec:MixedTrainEval}
Figure~\ref{fig:MixedTrainEval} demonstrate the inference accuracy curve after on-chip deployment using our mixed-training strategy.
Active phases are randomly selected from all phases.
With $\Delta\gamma\in\mathcal{N}(0, 0.002)$ phase shifter variations and $\omega$=2e-3 thermal crosstalk between adjacent devices, the initial accuracy varies among different mixed-training sparsity values.
A larger mixed-training sparsity can provide protection to the PIC from thermal crosstalk since more passive devices generate fewer thermal noises.
The convergence of those curves shows that only a small fraction (5\%-15\%) of devices is necessary to perform on-chip learning for accuracy recovery, while simply tuning every parameter has the lowest efficiency and effectiveness among all settings.
When $\alpha$ is set to 5\%-15\%, we observe the fastest convergence speed,
leading to 3.7$\times$-7.6$\times$ higher training efficiency, i.e., fewer function queries, than ours with a large $\alpha$.
This mixed-training strategy makes it possible to recover the accuracy of larger-scale ONNs with much fewer function queries and lower power.

\subsection{Evaluation on the Sparsity of \texttt{SZO-SCD}}
\label{sec:SparseTuneEval}
Figure~\ref{fig:SparseTuneEval} demonstrates how different sparsity $s$ influences the on-chip learning performance under a given mixed-training sparsity $\alpha=0.15$.
Among all sparsity values, 60\% is the best tuning sparsity that leads to the fastest convergence speed on small datasets.
In contrast, on MNIST, since the dataset variance is much larger than Vowel Recognition, a highly-sparse learning strategy ($s<$0.1) is more suitable to balance the variance and generalization in the stochastic optimization.
In other words, overly-greedy optimization caused by large $s$ values is harmful to stochastic learning.
Note that a higher sparsity, e.g., $s<$0.02, will lead to accuracy loss since the variance is too large for the optimizer to converge, which is not shown on the figure for brevity.

\subsection{Compare with Other Zeroth-Order Optimizers}
\label{sec:CompareOptimizer}
To validate the efficiency of our proposed \texttt{SZO-SCD}, we compare a variety of state-of-the-art ZO optimizers on different sparsity in Table.~\ref{tab:CompareOptimizer}.
Proper $\alpha$ and $s$ are adopted to obtain a good trade-off between accuracy and efficiency.
Learning rates reported are empirically most suitable values with equal parameter searching efforts for all methods.
The comparison results provide several important insights.
First, in high-dimensional ONN parameter space, the gradient-based methods, e.g., \texttt{FLOPS}, generally show poor performance and unstable convergence due to gradient estimation variance.
Even for \texttt{ZOO-ADAM} and \texttt{ZOO-Newton} that adaptively adjust the step size, they suffer from divergence in the phase-domain optimization unless a descent in the objective can be partially guaranteed like our proposed coordinate descent method.
Second, the two-level sparsity is indeed necessary to achieve stable convergence and good model generalization.
\texttt{FLOPS} with two-level sparsity coincides with \cite{NN_Arxiv2020_Ohta} and shows better convergence than the dense counterpart.
Third, gradient-based methods require an arbitrarily tiny step size ($<$1e-3) for gradient estimation and weight updating, which is not practical given limited device control resolution.
In contrast, our method only needs a medium step size, corresponding to 8-bit control precision, a more hardware-friendly configuration in analog neuromorphic computing.
Fourth, interestingly, the stochastic coordinate-wise three-points method (\texttt{STP}) leads to worse inference accuracy than our method due to its overly-greedy updating mechanism that potentially harms generalization as follows,
\begin{equation}
    \small
    \label{eq:STP}
    \phi_i^t \gets \argmin_{\phi_i}\{\mathcal{L}(\phi_i^{t-1}),\mathcal{L}(\phi_i^{t-1}+\delta\phi),\mathcal{L}(\phi_i^{t-1}-\delta\phi)\}.
\end{equation}
Table.~\ref{tab:CompareOptimizerAverage} further shows the average performance on different datasets with CNNs.
Overall, our proposed mixed-training strategy with \texttt{SZO-SCD} shows the best convergence and accuracy with the smallest computation and memory cost.

\subsection{Evaluation on the Power-Aware Dynamic Pruning}
\label{sec:PowerAwarePruneEval}
\begin{figure*}
    \centering
    \subfloat[]{\includegraphics[width=0.3\textwidth]{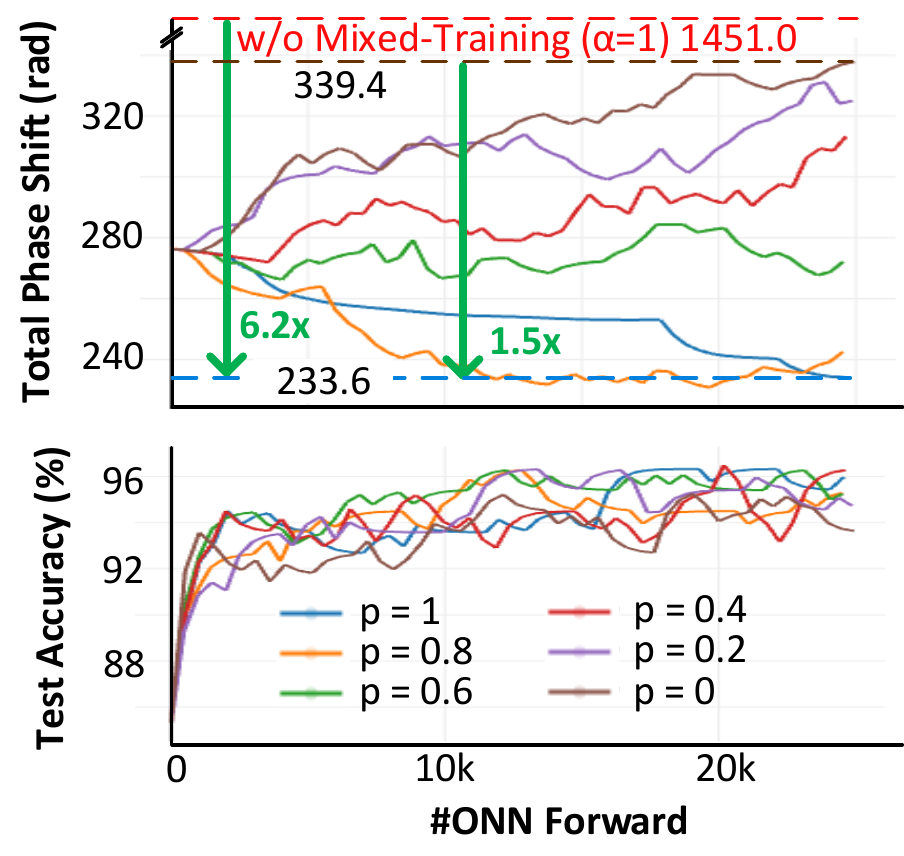}
    }
    \subfloat[]{\includegraphics[width=0.3\textwidth]{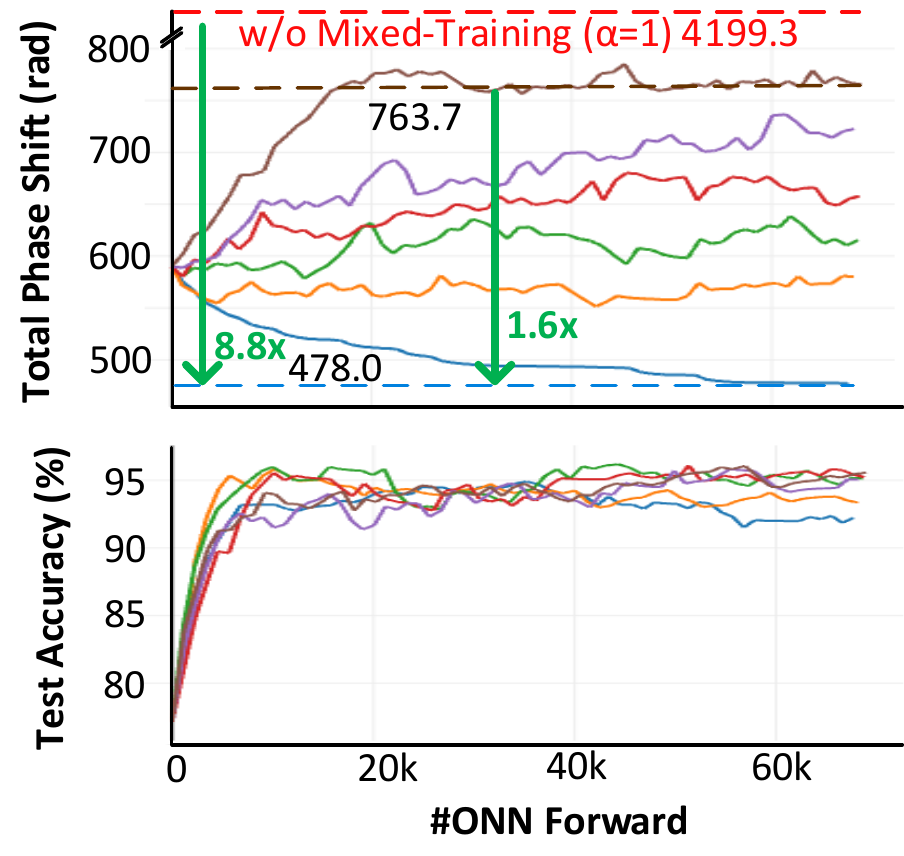}
    }
    \subfloat[]{\includegraphics[width=0.3\textwidth]{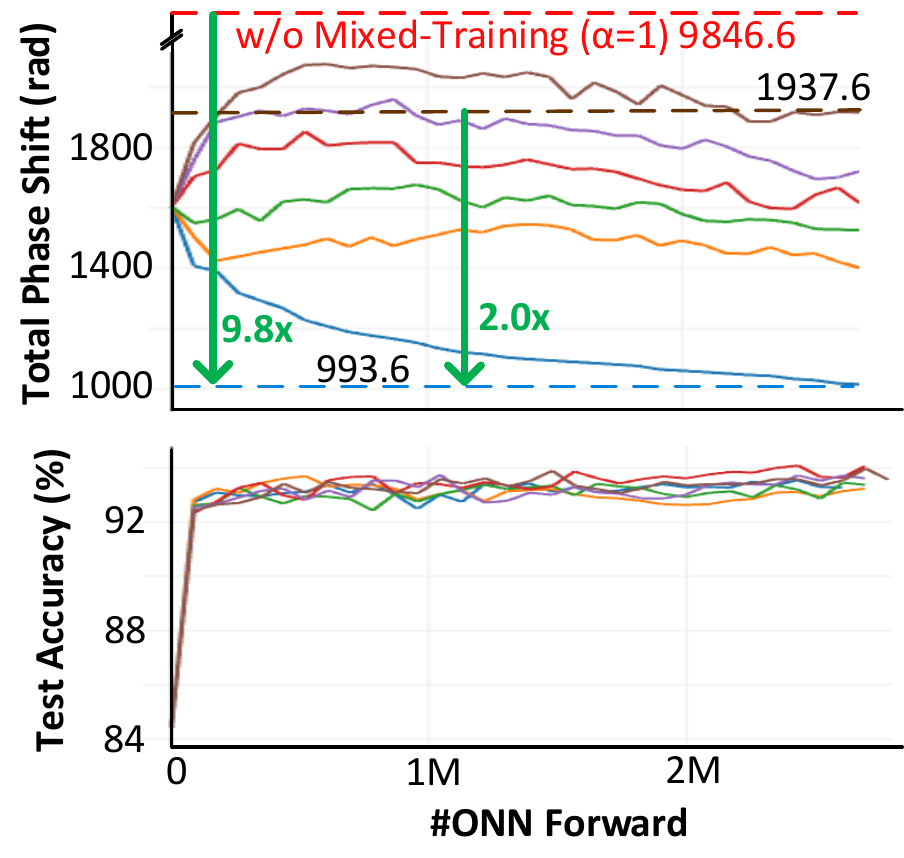}
    }
    \caption{Estimated power and inference accuracy evaluation with different power awareness $p$.
    The mixed-training sparsity $\alpha$ is selected as 0.15.
    The sparsity $s$ for \texttt{SZO-SCD} is set to 0.6 for (a) and (b), and is set to 0.1 for (c).
    The model and dataset are the same as Fig.~\ref{fig:MixedTrainEval}
    }
    \label{fig:PowerEval}
    \vspace{-5pt}
\end{figure*}

We evaluate the effectiveness of our proposed power-aware dynamic pruning technique in Fig.~\ref{fig:PowerEval}.
Different power awareness values lead to slightly different inference accuracy after convergence.
However, fully-power-aware ($p$=1) pruning can cut down 30\%-50\% power compared with the power-unaware version ($p$=0).
Compared with the naive ONN deployment without mixed-training, our power-aware mixed-training can save a total $\sim$90\% power.
This lightweight pruning method not only reduces the power in inference $\mathcal{P}$, shown in the final power value at the end of the curve, it also saves the training energy $\int_t\mathcal{P}\;\text{d}t$ indicated by the area under the power curve.
We also compare with ADMM to show the superiority of our dynamic pruning technique in Table~\ref{tab:ComparePowerOptimizer}.
The Lagrangian-relaxation-based formulation and ADMM-based optimization algorithm are not suitable for power-aware ONN on-chip learning.
A small $\lambda$ in the dual update step has a trivial influence on the total power, while a large $\lambda$ leads to unstable convergence.
In contrast, our proposed method can provide stable power constraint handling with a parametric mechanism to achieve a trade-off between accuracy and power.
\begin{table*}[t]
\centering
\caption{Comparison with SOTA ZO optimizers in terms of optimizer cost per iteration, ONN query complexity per iteration, and memory complexity.
\textit{lr} is the step size.
We evaluate on MNIST with a 3-layer optical MLP (64-24-24-10).
$T$ is the total iteration.
$d$ is the total number of variables ($d$=2,350).
The sampling factor $Q$ is set to 60 as used in \texttt{FLOPS}~\cite{NP_DAC2020_Gu}.
}
\resizebox{0.95\linewidth}{!}{%
\begin{tabular}{l|lllllll}
\hline\hline
Optimizer$\quad$ & $\alpha\quad\quad$ & $s\quad\quad$ & lr$\quad\quad$ & Computation$\quad$ & \#ONN forward$\quad$ & Memory$\quad$ & Best Acc. \\\hline
\texttt{ZOO-ADAM}~\cite{NN_AIsec2017_Chen}   & 1    & 1   & 1e-3 & $\mathcal{O}(d)$         & $2Td~(4700T)$           & $\mathcal{O}(d)$ &  diverge\\
\texttt{ZOO-ADAM}~\cite{NN_AIsec2017_Chen}   & 0.15 & 0.1 & 1e-3 & $\bm{\mathcal{O}(\alpha sd)}$ & $2T\alpha sd~(70.5T)$   & $\mathcal{O}(\alpha d)$ & 88.1\%\\
\texttt{ZOO-Newton}~\cite{NN_AIsec2017_Chen} & 1    & 1   & 1e-3 & $\mathcal{O}(d)$         & $3Td~(7050T)$ & $\bm{\mathcal{O}(1)}$ & diverge\\
\texttt{ZOO-Newton}~\cite{NN_AIsec2017_Chen} & 0.15 & 0.1 & 1e-3 & $\bm{\mathcal{O}(\alpha sd)}$ & $3T\alpha sd~(105.75T)$   & $\bm{\mathcal{O}(1)}$ & diverge\\
\texttt{STP}~\cite{NN_AAAI2020_Bibi} & 1    & 1   & 2e-2 & $\mathcal{O}(d)$         & $2Td~(4700T)$ & $\bm{\mathcal{O}(1)}$ & diverge\\
\texttt{STP}~\cite{NN_AAAI2020_Bibi} & 0.15 & 0.1 & 2e-2 & $\bm{\mathcal{O}(\alpha sd)}$ & $2T\alpha sd~(70.5T)$   & $\bm{\mathcal{O}(1)}$ & 90.2\%\\
\texttt{FLOPS}~\cite{NP_DAC2020_Gu}          & 1    & 1   & 1e-1 & $\mathcal{O}(Qd)$        & $TQ~(60T)$            & $\mathcal{O}(d)$ &  diverge\\
\texttt{FLOPS}~\cite{NP_DAC2020_Gu}          & 0.15 & 0.1 & 1e-1 & $\mathcal{O}(Qd)$        & $TQ~(60T)$            & $\mathcal{O}(\alpha sd)$ &  83.5\%\\
\texttt{SZO-SCD}~(Proposed)                  & 0.15 & 0.1 & 2e-2 & $\bm{\mathcal{O}(\alpha sd)}$ & $\bm{1.5T\alpha sd~(52.88T)}$ & $\bm{\mathcal{O}(1)}$ & \textbf{93.5\%}\\\hline\hline
\end{tabular}
}
\label{tab:CompareOptimizer}
\end{table*}

\begin{table}[]
\centering
\caption{Average accuracy(std.) among different optimizers over 3 runs.
The CNN setting is 16$\times$16-c8s2-c6s2-10 for MNIST,
32$\times$32-c8s2-c8s2-10 for FMNIST, and 32$\times$32-c8s2-c8s2-c8s2-10 for CIFAR-10.
\textit{c8s2} is 8 kernels with size 3$\times$3 and stride 2.
$\alpha$ and $s$ are set to 0.05 and 0.1 for all optimizers.
}
\resizebox{\linewidth}{!}{%
\begin{tabular}{l|lcc}
\hline\hline
Optimizer & MNIST & FMMIST & CIFAR-10 \\\hline\texttt{ZOO-Adam}   &88.51\%(0.10)      &  68.16\%(0.13)      &  diverge  \\
\texttt{ZOO-Newton}   & diverge      & 67.60\%(0.23)      &  diverge  \\
\texttt{STP}   & 93.74\%(0.30)      &  75.43\%(3.86)      &  diverge  \\
\texttt{FLOPS}   & diverge      &  67.27\%(0.19)      &  diverge  \\
\texttt{SZO-SCD}   & 94.88\%(0.26)      &  82.63\%(0.04)      &  51.35\%(0.86)  \\\hline\hline
\end{tabular}
}
\vspace{-5pt}
\label{tab:CompareOptimizerAverage}
\end{table}

\begin{table}[]
\centering
\caption{Comparison among ADMM-based power-constrained optimization and our proposed dynamic power-aware pruning.
Power is estimated by the total phase shifts of active MZIs.
$\lambda$ is the weight for the relaxed power penalty term.
The model configuration of the 3-layer optical MLP is 64-24-24-10, and the dataset is downsampled MNIST.
$\alpha$ and $s$ are set to 0.15 and 0.1 respectively.
}
\resizebox{\linewidth}{!}{%
\begin{tabular}{l|lcc}
\hline\hline
Method & Hyperparameter & Power (rad) & Test Accuracy \\\hline
ADMM   & $\lambda=$~0.05$\sim$0.3      &  2076.6$\sim$2367.1      &  92.8\%$\sim$79.0\%  \\
ADMM   & $\lambda>$~0.3      &  -      &  diverge  \\
Proposed   & $p=$~0$\sim$1      &  1937.6$\sim$993.6      &  93.9\%$\sim$93.1\%  \\\hline\hline
\end{tabular}
}
\vspace{-5pt}
\label{tab:ComparePowerOptimizer}
\end{table}
\subsection{Evaluation on CNNs and Different Datasets}
\label{sec:CNNEval}
We further evaluate the effectiveness of our proposed mixed-training strategy with sparse tuning on convolutional neural networks (CNNs).
We use \textit{im2col} algorithm to implement convolution with general matrix multiplication (GEMM).
Table~\ref{tab:EvalCNN} shows the accuracy recovery results and power improvement on three different datasets compared with w/o mixed-training or power handling.
On three practical datasets, our proposed methods demonstrate stable accuracy recovery for optical CNN architectures under device variations while reducing the total inference power by $>$95\%.
\begin{table}[]
\centering
\caption{Power reduction on different datasets with CNNs (same as Table.~\ref{tab:CompareOptimizerAverage}).
\textit{DAcc.} and \textit{RAcc.} represent deployed and recovered accuracy, respectively.
\textit{PR-Ours} and \textit{PR-FLOPS} are power reduction compared to ours($p$=0) and \texttt{FLOPS}.
}
\resizebox{\linewidth}{!}{%
\begin{tabular}{l|llllcccc}
\hline\hline
Dataset  & $\alpha$ & $s$ & $p$ & DAcc. & RAcc. & PR-Ours. & PR-\texttt{FLOPS}\\\hline
MNIST    & 0.05 & 0.1 & 1 & 87.4\% & 95.5\% & 98.8\% & 97.6\%\\
FMNIST   & 0.05 & 0.1 & 1 & 65.7\% & 82.6\% & 95.6\%  & 98.1\%\\
CIFAR-10 & 0.05 & 0.1 & 1 & 36.0\% & 52.5\% & 96.7\% & 96.7\% \\\hline\hline
\end{tabular}
}
\vspace{-5pt}
\label{tab:EvalCNN}
\end{table}

\vspace{-.05in}
\section{Conclusion}
\label{sec:Conclusion}
In this work, we propose a scalable ONN on-chip learning framework to efficiently perform \textit{in situ} accuracy recovery with dynamic power optimization.
We are the first to formulate ONN on-chip learning problem with device non-ideality and power constraints.
A mixed-training strategy with sparse coordinate descent \texttt{SZO-SCD} is proposed to explore two-level sparsity in ONN deployment and optimization, leading to better training efficiency and robustness.
A lightweight dynamic power-aware pruning is proposed to implicitly optimize power during \textit{in situ} learning with near-zero computational cost or accuracy loss.
Compared with SOTA ONN on-chip learning methods, our proposed framework boosts the efficiency by 3.7$\times$-7.6$\times$ with better crosstalk-robustness, 2$\times$ better scalability, and over 10$\times$ better power efficiency.

\section*{Acknowledgment}
The authors acknowledge the Multidisciplinary University Research Initiative (MURI) program through the Air Force Office of Scientific Research (AFOSR), contract No. FA 9550-17-1-0071, monitored by Dr. Gernot S. Pomrenke.

\end{document}